**Cyclic Coding Algorithms under MorphoSys Reconfigurable Computing System**

**Hassan Diab**
diab@aub.edu.lb
Department of ECE
Faculty of Eng'g and Architecture
American University of Beirut
P.O. Box 11-0236
Beirut, Lebanon

**Issam Damaj**
damajiw@sbu.ac.uk
Faculty of Eng'g, Science and Technology
South Bank University
103 Borough RD
SE1 0AA
London, U.K.

**ABSTRACT:** This paper introduces reconfigurable computing (RC) and specifically chooses one of the prototypes in this field, MorphoSys (M1) [1 - 5]. The paper addresses the results obtained when using RC in mapping algorithms pertaining to digital coding in relation to previous research [6 - 10]. The chosen algorithms relate to cyclic coding techniques, namely the CCITT CRC-16 and the CRC-16. A performance analysis study of the M1 RC system is also presented to evaluate the efficiency of the algorithm execution on the M1 system. For comparison purposes, three other systems where used to map the same algorithms showing the advantages and disadvantages of each compared with the M1 system. The algorithms were run on the 8x8 RC (reconfigurable) array of the M1 (MorphoSys) system; numerical examples were simulated to validate our results, using the MorphoSys mULATE program, which simulates MorphoSys operations.

## 1. INTRODUCTION

Reconfigurable computing (RC) is becoming more popular and increasing research efforts are being invested in it [1-10]. It employs reconfigurable hardware and programmable processors. The user designs the program in a way where the workload is divided between the general-purpose processor and the reconfigurable device. The use of RC paves the way for an increased speed over general-purpose processors and a wider functionality than Application Specific Integrated Circuits (ASICs). It is a solution for applications requiring a wide range of functionality and speed [1]. RC Systems represent a solution to the inflexibility of ASICs on one end of the computing spectrum, and the inefficiency of General Purpose Processors (GPPs) on the other end of the spectrum. Reconfigurable computers (RCs) offer the potential to greatly accelerate the execution of a wide variety of applications. Its key feature is the ability to perform computations in hardware to increase performance, while retaining much of the flexibility of a software solution.

## 2. MORPHOSYS DESIGN

One of the emerging RC systems includes the MorphoSys designed and implemented at the University of California, Irvine. It has the block diagram shown in Figure 1 [2] and is composed of: 1) an array of reconfigurable cells called the RC array, 2) its configuration data memory called context memory, 3) a control processor (TinyRISC), 4) a data buffer called the frame buffer, and 5) a DMA controller [2]. A program runs on MorphoSys in the following manner: General-purpose operations are handled by the TinyRISC processor, while operations that have a certain degree of parallelism, regularity, or intensive computations are mapped to the RC array.

The TinyRISC processor controls, through the DMA controller, the loading of the context words to context memory. These context words define the function and connectivity of the cells in the RC array. The processor also initiates the loading of application data, such as image frames, from main memory to the frame buffer. This is also done through the DMA controller. Now that both configuration and application data are ready, the TinyRISC processor instructs the RC array to start execution. The RC array performs the needed operation on the application data and writes it back to the frame buffer. The RC array loads new application data from the frame buffer and possibly new configuration data from context memory. Since the frame buffer is divided into two sets, new application data can be loaded into it without interrupting the operation of the RC array. Configuration data is also loaded into context memory without interrupting RC array operation. This causes MorphoSys to achieve high speeds of execution [3].

## 3. RECONFIGURABLE DEVICE

As stated earlier, the reconfigurable device in MorphoSys is the RC array divided into four quadrants. It has the design and interconnection shown in Figure 2 [2]. The interconnection network is built on three hierarchical levels. The first is a nearest neighbor layer that connects the RCs in a 2-D mesh. The second is an intra-quadrant connection that connects a specific RC to any other RC in its row or column in the same quadrant. The third is an inter-quadrant connection that carries data from any one cell (out of four) in a row (or column) of a quadrant to other cells in an adjacent quadrant but in the same row (or column) [4].

The context words loaded into context memory configure the function of the RCs as well as the interconnection, thus specifying where their input is from and where their output will be written to [5]. MorphoSys is designed in a way where all the cells in the same row perform the same function and have the same connection scheme (in row context broadcast mode), or all the cells in the same column perform the same function and have the same connection scheme (in column context broadcast mode). All the cells of a row or of a column share the same configuration [5].

## 4. CODING ALGORITHMS UNDER MORPHOSYS

Reconfigurable hardware implementation of digital coding algorithms has been an active area of research [6 - 10]. Many coding algorithms where mapped onto the M1. Research done to date includes: performance study of coding algorithms (checksum), pipelined implementation of various non-standard linear sequential coding circuits, and other algorithms [8,10]. The linear sequential circuits considered here are finite state machines with a finite number of inputs and outputs. The inputs, outputs and state transition occur at discrete intervals of time. The elements used are adders (EX-OR) and the delays (D) to delay input words. A sequence of 0s and 1s can be expressed by a polynomial in which the 0s and 1s are coefficients of the powers of a dummy variable. Hence, the sequence 11001 can be written as:

$$1D^4 \oplus 1D^3 \oplus 0D^2 \oplus 0D^1 \oplus 1D^0$$

This representation is the basis of the Feed Forward Binary Circuits, which are very useful in coding techniques. A generalized form, shown in Figure 3, of these circuits is represented by

$$T(D) = (1 \oplus D^1 \oplus D^2 \oplus D^3)$$

This circuit is used to code any stream of input vector X and yields a set of outputs Y. Therefore, the input vector X is a vector of 0s and 1s and the output is the coded output Y vector, which is the result of multiplying the input polynomial (vector) X with the polynomial represented by T(D). This could be generalized to take the form:

$$y_k = \sum_{j=0}^{N} \otimes x_{k-j} D^j ,$$

or, finally

$$Y = (\sum_{j=0}^{N} \otimes D^j)X \text{ i.e. } Y = (D^0 \oplus D^1 \oplus D^2 \oplus ... \oplus D^N)X$$

Where, N is the number of stages of the circuit, and X is of the form $X = x_0...x_{k-1}x_k$ with $x_k$ being the first bit to enter the multiplier circuit. This paper focuses on the hardware implementation of cyclic redundancy codes checkers (CRCCs) with their standard circuits, namely the CRC-16 circuits.

## 5. CYCLIC REDUNDANCY CODES

Redundant encoding is a method of error detection that spreads the information across more bits than the original data. The more redundant bits used, the greater the chance to detect errors. CRCCs are check for differences between transmitted data and the original data. CRCCs are effective for two reasons: Firstly, they provide excellent protection against common errors, such as burst errors where consecutive bits in a data stream are corrupted during transmission. Secondly, systems that use CRCCs are easy to implement [11]. When a CRCC is used to verify a frame of data, the frame is treated as one very large binary number, which is then divided by a generator number. This division produces a reminder, which is transmitted along with the data. At the receiving end, the data is divided by the same generator number and the remainder is compared with the one sent at the end of the data frame. If the two remainders are different, then an error occurred during data transmission. Types of errors that a CRCC detects depend on the generator polynomial. Table 1 shows the most common generator polynomials.

### 5.1. CRC Serial Implementation

CRC implementation is usually done with linear-feedback shift registers (LFSRs). Figures 4 and 5 show the CCITT CRC-16 and CRC-16 generators with their serial implementation using LFSR. This serial method works well when the data is available in bit-stream form.

### 5.2. CRC Parallel Implementation

With the currently available high-speed digital signal processing (DSP) systems, the processing of data is done in a byte, word, double word, or larger widths rather than serially. Even with serial telecommunication systems, data is buffered in chips responsible for synchronizations and framing. For parallel implementation the data is available in 8-bit frames with manageable speed [11]. A one channel parallel CRC algorithm with LFSR approach is done by considering the state of the circuit on 8-shifts basis [11 - 12]. Tables 2-3 show two different implementations of the CCITT CRC-16 and CRC-16, respectively. The term $Register_i$ represents the LFSR internal register numbered "i", while $XOR_j$ represents the output of the $XOR$ gate number "j", and $XOR$ indicates the XOR operation. With the emergence of the highly scalable reconfigurable circuits, more implementation capabilities are present. Along with the byte-wise or word-wise CRC implementation it is possible to implement parallel channels each with byte-wise CRC implementation.

# 6. ALGORITHMS MAPPING

From the underlying architecture point of view, the mapping of any algorithm onto the proposed reconfigurable system requires in-depth knowledge of all the available interconnection topologies. Moreover, the designer should take into consideration the possibility of dynamically changing the shape of the interconnection. From the algorithmic point of view, the design of a parallel version of the addressed algorithms requires the best use of recourses with the least possible redundancy in computations.

Three sets of data are required to map any algorithm onto the M1 system. The first set specifies the intended shapes of interconnections that are going to be used. The second set is the manipulated data. Lastly, the last set of code is the TinyRISC code that will orchestrate the load/save operations, parallel computations, and changing the interconnection pattern through the context words.

## 6.1. The CCITT CRC-16 Algorithm Mapping

The mapping of the parallel CCITT CRC-16 algorithm will make use of the redundant computations utilized in several steps of the algorithm. Firstly, the values of $XOR_i$ for all values of i from 0 to 7 are calculated. In Table 2 the computations that are used more than once are shown. Particularly, the redundant values are $(XOR_{i+4} \oplus XOR_i)$ for i from 0 to 3. Thus, the second computation step involve registers 4, 5, 6, 11, 12, 13, 14, and 15 depending on results found in the first step. In the final step the computations for registers 0,1,2,3,7,8,9 and 10 are carried out depending on the results calculated in the first and second step.

The algorithm mapping will be explained by introducing the three needed sets of code. The first set is the interconnection context words. The context word used in this algorithm is that for XOR with column broadcast, where each cell XORs two inputs from frame buffers A and B. This context word is stored at address $30000_{hex}$.

The second set of code is the input data and the initial data in the circuits registers. These two sets of data are stored in main memory address $10000_{hex}$ and $20000_{hex.}$

The third set is that of the TinyRISC code, which is the main code. This code and its discussion are shown in Table 4. Main steps of the addressed algorithm are shown in Figures 6 and 7. The final contents of frame buffer A is shown in Figure 8.

## 6.2. The CRC-16 Algorithm Mapping

The mapping of this algorithm depends also on eliminating redundant computations, besides, the parallel computation of the required values. This mapping is of three steps. Firstly, the values of $XOR_i$ for all values of i from 0 to 7 are calculated. From Table 3 the computations used more than once are shaded, particularly, the redundant value X $(XOR_0 \oplus ... \oplus XOR_7)$. Thus, the second computation step is for X. Thirdly, the rest of the values are calculated in parallel.

The algorithm mapping will be explained by introducing the three needed sets of code. The first set is the interconnection context words. The context words used in this algorithm are firstly, that for XOR with column broadcast, where each cell XORs two inputs from frame buffers A and B. Secondly, the same cell operation is used also by taking one input from the frame buffer, and the second from the output of the left adjacent cell. The context words are stored at address $30000_{hex}$. The second set of code is the input data and the initial data in the circuits registers. These two sets of data are stored in main memory address $10000_{hex}$ and $20000_{hex}$. The third set is that of the TinyRISC code which is the main code, this code and its discussion are shown in Table 5. Main steps of the addressed algorithm are shown in Figures 9 and 10. The final contents of frame buffer A is shown in Figure 11.

## 7. PERFORMANCE EVALUATION AND ANALYSIS

The performance is based on the execution speed of the algorithms presented in sections 6.1 and 6.2 corresponding to Tables 2 and 3 respectively, which show the states of the registers after 8 shifts. The MorphoSys system is considered to be operational at a frequency of 100 MHz.

The algorithm in Table 4 (CRC-CCITT-16 Parallel Algorithm for a single channel) takes 30 cycles to complete. The speed in bits per cycle of the algorithm of Table 4 is equal to 0.267 bits/cycles i.e. 3.75 cycles for each bit. The time for the algorithm to terminate is equal to 0.3 μsec, and the data rate is 26.67 Mbps.

The algorithm in Table 5 (CRC-16 Parallel Algorithm for a single channel) takes 26 cycles in order to terminate. The cycle time for the MorphoSys is equal to 10 nsec. Thus, the speed in bits per cycle of the algorithm of Table 5 is equal to 0.307 bits/cycles i.e. 3.25 cycles for each bit. The time for the algorithm to terminate is equal to 0.26 μsec, and then the rate in Mega bits per second (Mbps) is 30.76 Mbps.

Furthermore, a comparison is done with the same algorithms mapped onto some Intel microprocessing systems. In this research the chosen processors are the Intel 80486 and Pentium. Note that the instructions used are upward compatible with newer Intel processors. The code and discussion of the same algorithms in Tables 4 and 5 are shown in Tables 6 and 7 respectively. Note that the chosen systems have comparable frequencies of 100 ~ 133 MHz.

In addition to Intel systems, the RC-1000 FPGA is used for performance comparisons. The Celoxica RC1000 board provides high-performance, real-time processing capabilities and is optimized for the Celoxica DK1 design suite. The RC1000 is a standard PCI bus card equipped with a Xilinx® Virtex™ family BG560 part with up to 2 million system gates. It has 8MB of SRAM directly connected to the FPGA in four 32-bit wide memory banks. The memory is also visible to the host CPU across the PCI bus as if it were normal memory. Each of the 4 banks may be granted to either the host SRAM on the board. It is then accessible to the FPGA directly and to the host CPU either by DMA transfers across the PCI bus or simply as a virtual address. Comparisons among those systems are shown in Tables 8 and 9.

For the maximum exploitation of the M1 capabilities, it should be noted that the M1 data items are byte-wise (8 bits). Thus, the M1 can calculate in parallel the input of up to 8-channels simultaneously. This is also shown in Tables 8 and 9. Note that the FPGA RC-1000 findings are the same for a single channel or 8-channels input because of its scalability. The speedup factors, besides the other chosen metrics, show the superiority of the used reconfigurable computing systems. The speedup factor is considered to be the ratio between the cycle times of the suggested systems.

## 8. CONCLUSION

New mapping algorithms are introduced dealing with coding operations and its performance analysis under MorphoSys is proposed. Many findings besides the speed of these mappings are calculated, and results are compared with other processing systems. The cyclic coding algorithms are presented with their mapping onto the M1. Accordingly, speeds of 213.13 Mbps for the CRC-CCITT-16 and 246.15 Mbps for the CRC-16 were achieved. The speedup factors (ratio of number of clocks) ranged from 4.26 to 58.46 between the M1 and the Intel processing systems. Moreover, the speed up factors between the RC-1000 and the M1 were up to 3.75. Future efforts could be invested in trying to map other algorithms that make use of the already mapped ones for more advanced algorithms for digital coding. The current research includes the work with other cyclic redundancy check algorithms along with other state-of-the-art coding methods. Also, comparisons could be made with results available on other parallel processors.

## Biographies

### Biography for Hassan Diab

**Hassan Diab** received his B.Sc. in Communications Engineering, M.Sc. in Systems Engineering, and Ph.D. in Computer Engineering. He is a Professor of Electrical and Computer Engineering at the American University of Beirut, Lebanon. He has over 90 publications in international journals and conferences. His research interests include performance evaluation of parallel processing systems, reconfigurable computing, and simulation of parallel applications. Professor Diab is a Fellow of IEE and IEAust and a Senior Member of IEEE.

### Biography for Issam Damaj

**Issam Damaj** received his B.Eng. in Computer Engineering, M.Eng. in Computer and Communications Engineering, and is currently a Ph.D. student at South Bank University, London. He is a Teaching Assistant in the Faculty of Engineering, Science and Technology at South Bank University, London. His research interests include reconfigurable computing, H.W./S.W. Co-Design, fuzzy logic, and wireless communications security. He is a Member of the IEEE.

## List of Figures:

**<u>List of Tables:</u>**

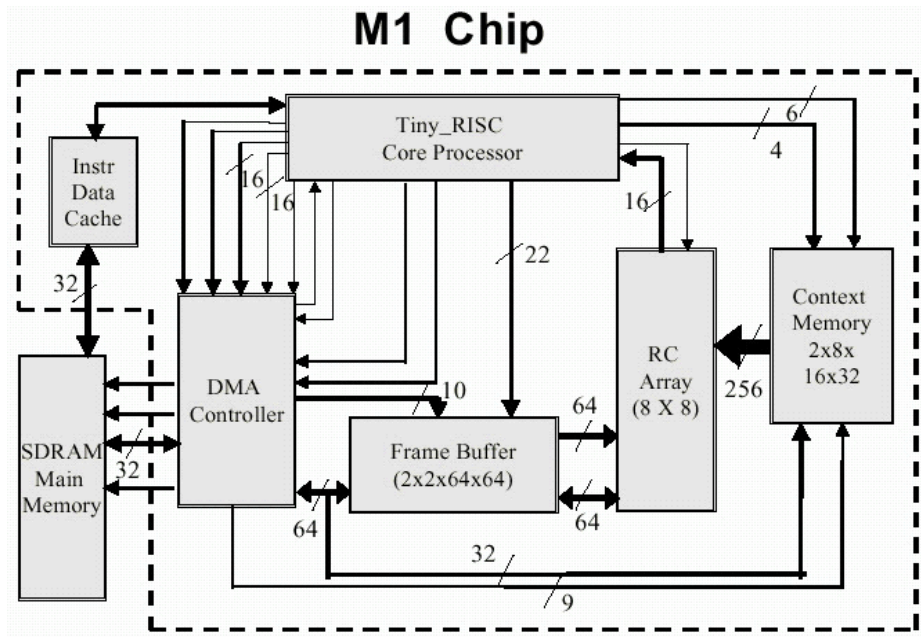

Figure 1. MorphoSys Block Diagram.

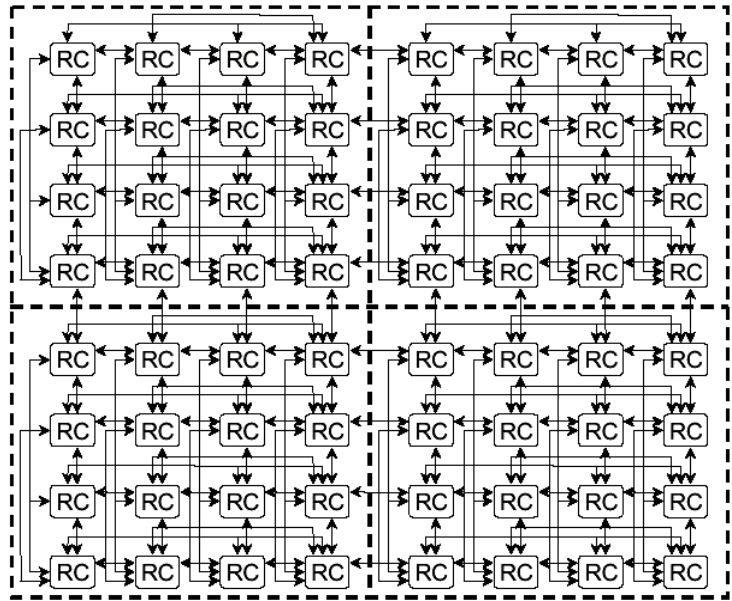

Figure 2.  RC Array Interconnection.

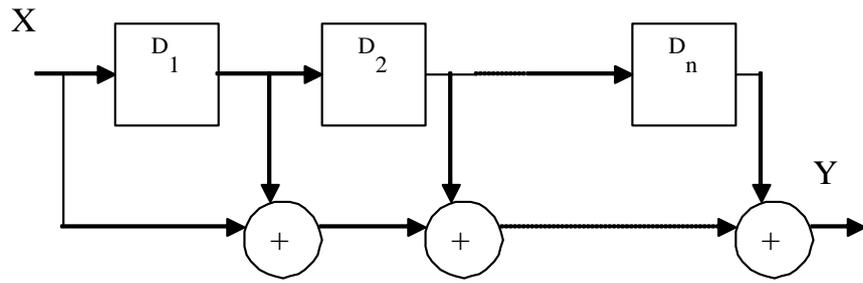

Figure 3. A General Representation of Feed Forward Binary Circuits.

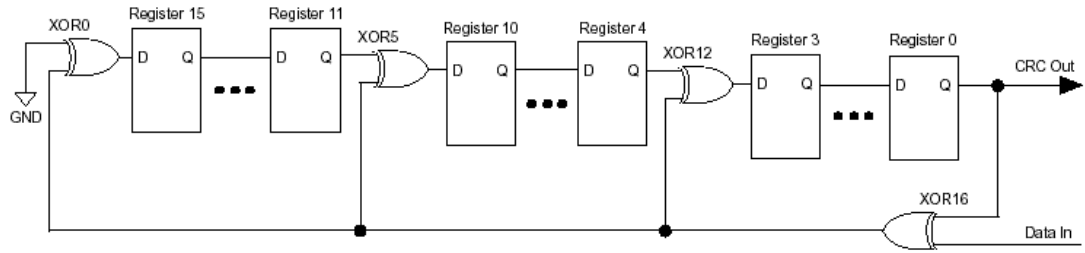

Figure 4. LFSR implementation of the CCITT CRC-16

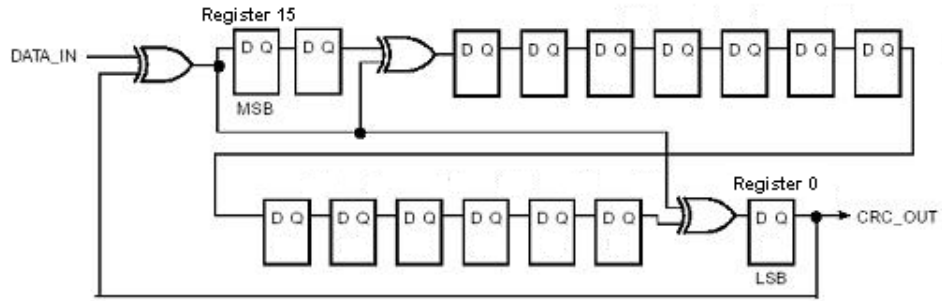

Figure 5. LFSR implementation of the CRC-16

| Columns Rows | $C_0$ | $C_1$ | $C_2$ | $C_3$ | $C_4$ | $C_5$ | $C_6$ | $C_7$ |
|---|---|---|---|---|---|---|---|---|
| $R_0$ | $Register_0 \oplus DataIn_0$ | . | . | . | . | . | . | . |
| $R_1$ | $Register_1 \oplus DataIn_1$ | . | . | . | . | . | . | . |
| $R_2$ | $Register_2 \oplus DataIn_2$ | . | . | . | . | . | . | . |
| $R_3$ | $Register_3 \oplus DataIn_3$ | . | . | . | . | . | . | . |
| $R_4$ | $Register_4 \oplus DataIn_4$ | . | . | . | . | . | . | . |
| $R_5$ | $Register_5 \oplus DataIn_5$ | . | . | . | . | . | . | . |
| $R_6$ | $Register_6 \oplus DataIn_6$ | . | . | . | . | . | . | . |
| $R_7$ | $Register_7 \oplus DataIn_7$ | . | . | . | . | . | . | . |

Figure 6. RC array contents after calculating for $XOR_i$.

| Columns<br>Rows | $C_0$ | $C_1$ | $C_2$ | $C_3$ | $C_4$ | $C_5$ | $C_6$ | $C_7$ |
|---|---|---|---|---|---|---|---|---|
| $R_0$ | $XOR_4 \oplus XOR_0$ | . | . | . | . | . | . | . |
| $R_1$ | $XOR_5 \oplus XOR_1$ | . | . | . | . | . | . | . |
| $R_2$ | $XOR_6 \oplus XOR_2$ | . | . | . | . | . | . | . |
| $R_3$ | $XOR_7 \oplus XOR_3$ | . | . | . | . | . | . | . |
| $R_4$ | . | . | . | . | . | . | . | . |
| $R_5$ | . | . | . | . | . | . | . | . |
| $R_6$ | . | . | . | . | . | . | . | . |
| $R_7$ | . | . | . | . | . | . | . | . |

Figure 7. RC array contents for the second computation step for CRC-CCITT-16.

| Address In HEX | Frame Buffer A | Address In HEX | Frame Buffer A |
|---|---|---|---|
| 0 | $DataIn_0$ | 40 | $\textbf{Register}_0$ |
| 1 | $DataIn_1$ | 41 | $\textbf{Register}_1$ |
| 2 | $DataIn_2$ | 42 | $\textbf{Register}_2$ |
| 3 | $DataIn_3$ | . | . |
| 4 | $DataIn_4$ | 50 | $\textbf{Register}_4$ |
| 5 | $DataIn_5$ | 51 | $\textbf{Register}_5$ |
| 6 | $DataIn_6$ | 52 | $\textbf{Register}_6$ |
| 7 | $DataIn_7$ | . | . |
| . | . | 55 | $\textbf{Register}_8$ |
| 10 | $XOR_0$ | 56 | $\textbf{Register}_9$ |
| 11 | $XOR_1$ | 57 | $\textbf{Register}_{10}$ |
| 12 | $XOR_2$ | . | . |
| 13 | $\textbf{XOR}_3 \textbf{ / Register}_{11}$ | 60 | $\textbf{Register}_3$ |
| 14 | $XOR_4$ | . | . |
| 15 | $XOR_5$ | 65 | $\textbf{Register}_7$ |
| 16 | $XOR_6$ | . | . |
| 17 | $XOR_7$ | . | . |
| . | . | . | . |
| 30 | $\textbf{XOR}_4 \oplus \textbf{XOR}_0 \textbf{/ Register}_{12}$ | . | . |
| 31 | $\textbf{XOR}_5 \oplus \textbf{XOR}_1 \textbf{/ Register}_{13}$ | . | . |
| 32 | $\textbf{XOR}_6 \oplus \textbf{XOR}_2 \textbf{/ Register}_{14}$ | . | . |
| 33 | $\textbf{XOR}_7 \oplus \textbf{XOR}_3 \textbf{/ Register}_{15}$ | . | . |

Figure 8. Contents of Frame Buffer A after the algorithm terminates after one computation step the new registers values are shown at the specified locations.

| Columns Rows | $C_0$ | ... | $C_7$ |
|---|---|---|---|
| $R_0$ | $Register_0 \oplus DataIn_0$ | . | . |
| $R_1$ | $Register_1 \oplus DataIn_1$ | . | . |
| $R_2$ | $Register_2 \oplus DataIn_2$ | . | . |
| $R_3$ | $Register_3 \oplus DataIn_3$ | . | . |
| $R_4$ | $Register_4 \oplus DataIn_4$ | . | . |
| $R_5$ | $Register_5 \oplus DataIn_5$ | . | . |
| $R_6$ | $Register_6 \oplus DataIn_6$ | . | . |
| $R_7$ | $Register_7 \oplus DataIn_7$ | . | . |

Figure 9. RC array contents after calculating for $XOR_i$.

| Columns<br>Rows | $C_0$ | $C_1$ | $C_2$ | $C_3$ | $C_4$ | $C_5$ | $C_6$ | $C_7$ |
|---|---|---|---|---|---|---|---|---|
| $R_0$ | $XOR_0$ | $XOR_0 \oplus XOR_1$ | . | . | . | . | . | $XOR_0 \oplus \ldots \oplus XOR_7$ |
| . | . | . | . | . | . | . | . | . |
| $R_7$ | . | . | . | . | . | . | . | . |

Figure 10. RC array contents for the second computation step for CRC-16.

| Address In HEX | Frame Buffer A | Address In HEX | Frame Buffer A |
|---|---|---|---|
| 0 | $DataIn_0$ | . | . |
| 1 | $DataIn_1$ | 36 | **$Register_8$** |
| 2 | $DataIn_2$ | 37 | **$Register_9$** |
| 3 | $DataIn_3$ | 38 | **$Register_{10}$** |
| 4 | $DataIn_4$ | 39 | **$Register_{11}$** |
| 5 | $DataIn_5$ | 40 | **$Register_{12}$** |
| 6 | $DataIn_6$ | 41 | **$Register_{13}$** |
| 7 | $DataIn_7$ | . | . |
| . | . | 45 | **$Register_0$** |
| 10 | $XOR_0$ | . | . |
| 11 | $XOR_1$ | 50 | **$Register_6$** |
| 12 | $XOR_2$ | . | . |
| 13 | $XOR_3$ | 55 | **$Register_7$** |
| 14 | $XOR_4$ | . | . |
| 15 | $XOR_5$ | . | . |
| 16 | $XOR_6$ | . | . |
| 17 | $XOR_7$ | . | . |
| . | . | . | . |
| 30 | **$XOR_0 \oplus \ldots \oplus XOR_7/Register_{14}$** | . | . |
| 31 | **$XOR_0 \oplus \ldots \oplus XOR_8/Register_{15}$** | . | . |

Figure 11. Contents of Frame Buffer A after the algorithm terminates after one computation step the new registers values are shown at the specified locations.

| Generator Name | Polynomial |
|---|---|
| SDLC (CCITT) | $X^{16} + X^{12} + X^5 + X^0$ |
| SDLC Reverse | $X^{16} + X^{11} + X^4 + X^0$ |
| CRC-16 | $X^{16} + X^{15} + X^2 + X^0$ |
| CRC-16 Reverse | $X^{16} + X^{14} + X^1 + X^0$ |
| CRC-12 | $X^{12} + X^{11} + X^3 + X^2 + X^1 + X^0$ |
| Ethernet | $X^{32} + X^{26} + X^{23} + X^{22} + X^{16} + X^{12} + X^{11} + X^{10} + X^8 + X^7 + X^5 + X^4 + X^2 + X^1 + X^0$ |

Table 1. Common generator polynomials.

| New Values After 8-shifts of the registers and the output of the XOR-gates | |
|---|---|
| $XOR_i =$ | $Register_i \oplus DataIn_i$       $I = 0, 1, \ldots, 7$ |
| | |
| $Register_0 =$ | $Register_8 \oplus XOR_4 \oplus XOR_0$ |
| $Register_1 =$ | $Register_9 \oplus XOR_5 \oplus XOR_1$ |
| $Register_2 =$ | $Register_{10} \oplus XOR_6 \oplus XOR_2$ |
| $Register_3 =$ | $Register_{11} \oplus XOR_0 \oplus XOR_7 \oplus XOR_3$ |
| $Register_4 =$ | $Register_{12} \oplus XOR_1$ |
| $Register_5 =$ | $Register_{13} \oplus XOR_2$ |
| $Register_6 =$ | $Register_{14} \oplus XOR_3$ |
| $Register_7 =$ | $Register_{15} \oplus XOR_4 \oplus XOR_0$ |
| $Register_8 =$ | $XOR_0 \oplus XOR_5 \oplus XOR_1$ |
| $Register_9 =$ | $XOR_1 \oplus XOR_6 \oplus XOR_2$ |
| $Register_{10} =$ | $XOR_2 \oplus XOR_7 \oplus XOR_3$ |
| $Register_{11} =$ | $XOR_3$ |
| $Register_{12} =$ | $XOR_4 \oplus XOR_0$ |
| $Register_{13} =$ | $XOR_5 \oplus XOR_1$ |
| $Register_{14} =$ | $XOR_6 \oplus XOR_2$ |
| $Register_{15} =$ | $XOR_7 \oplus XOR_3$ |

Table 2. The states of the registers after 8-shifts for the CCITT CRC-16 Algorithm.

| New Values After 8-shifts of the registers and the output of the XOR-gates | |
|---|---|
| $XOR_i =$ | $Register_i \oplus DataIn_i$      $i = 0, 1, …, 7$ |
| | |
| X | $XOR_0 \oplus XOR_1 \oplus … XOR_7$ |
| | |
| $Register_0 =$ | $Register_8 \oplus \overline{X}$ |
| $Register_1 =$ | $Register_9$ |
| $Register_2 =$ | $Register_{10}$ |
| $Register_3 =$ | $Register_{11}$ |
| $Register_4 =$ | $Register_{12}$ |
| $Register_5 =$ | $Register_{13}$ |
| $Register_6 =$ | $Register_{14} \oplus XOR_0$ |
| $Register_7 =$ | $Register_{15} \oplus XOR_1 \oplus XOR_0$ |
| $Register_8 =$ | $XOR_3 \oplus XOR_2$ |
| $Register_9 =$ | $XOR_4 \oplus XOR_3$ |
| $Register_{10} =$ | $XOR_5 \oplus XOR_4$ |
| $Register_{11} =$ | $XOR_6 \oplus XOR_5$ |
| $Register_{12} =$ | $XOR_7 \oplus XOR_6$ |
| $Register_{13} =$ | $XOR_8 \oplus XOR_7$ |
| $Register_{14} =$ | $XOR_8 \oplus \overline{X}$ |
| $Register_{15} =$ | $\overline{X}$ |

Table 3. The states of the registers after 8-shifts for the CRC-16 Algorithm.

| 0: | Ldui | r1, 0x1; | R1 ← 10000$_{hex}$. This is where DataIn Stored |
|---|---|---|---|
| 1: | Ldfb | r1, 0, 0, 2 ; | FB ← 2 x 32 bits at set 0, bank A, address 0. |
| 2: | Add | r0, r0, r0; | No-operation. |
| . | . | . | . |
| 6: | Ldui | r2, 0x1; | R2 ← 20000$_{hex}$. This is where Circuit Registers Stored. |
| 7: | ldfb | r2, 1, 0, 2 ; | FB ← 2 x 32 bits at set 0, bank B, address 0. |
| 8: | add | r0, r0, r0; | No-operation. |
| . | . | . | |
| 12: | ldui | r3, 0x3; | R3 ← 30000$_{hex}$. This is where the context word is stored in main memory. |
| 7: | ldctxt | r3, 0, 0, 0, 1; | Load 1 context word from main memory starting at the address stored in register 3 into plane 0, block 0 and starting at word 0. |
| 8: | add | r0, r0, r0; | NOP |
| . | . | . | |
| 11: | ldui | r4, 0x0; | R4 ← 00000hex. |
| 12: | dbcdc | r4, 0, 0, 0, 0, 0; | Double bank column broadcast. It sends data from both banks address 0 in the FB and broadcasts the context words column-wise. It triggers the RC array to start execution of column 0 by the context word of address 0 in the column block of context memory operating on data in set 0. Bank A starting at 0. Bank B starting at (0 + 0). |
| 13: | wfbi | 0, 0, 0, 0, 0x0; | Write data back to the FB A from the output registers of column 0 into set 0, address 0. Results for XOR$_i$. The value of Register$_{11}$. |
| 14: | wfbi | 0, 0, 1, 0, 10$_{hex}$; | Write data back to FB B from the output registers of column 0 into set 0, address 10$_{hex}$. |
| 15: | dbcdc | r4, 14$_{hex}$, 0, 0, 0, 1, 0; | Double bank column broadcast. Bank A starting at 0x0. Bank B starting at (0x0 + 14), i.e. shifted 4 words from the starting point of XOR$_0$. The calculated items in this operation are the repeatedly used operations shaded in Table 1. |
| 16: | wfbi | 0, 0, 0, 0, 30$_{hex}$; | Write data back to the FB from the output registers of column 0 into set 0, address 1D$_{hex}$. Values for Circuits Registers [12..15] are now available in FB A starting from address 30$_{hex}$. |
| 17: | dbcdc | r4, 8$_{hex}$, 0, 0, 0, 30$_{hex}$; | Double bank column 0 broadcast. Bank A starting at 30$_{hex}$. Bank B starting at (0x0 + 8$_{hex}$). Thus, Registers [0..2] values are now available. |
| 18: | dbcdc | r4, C$_{hex}$, 0, 1, 0, 1$_{hex}$; | Double bank column 1 broadcast. Bank A starting at 1$_{hex}$. Bank B starting at (0x0 + C$_{hex}$). Thus, Registers [4..6] values are now available. |
| 19: | dbcdc | r4, 10$_{hex}$, 0, 2, 0, 31$_{hex}$; | Double bank column 2 broadcast. Bank A starting at 30$_{hex}$. Bank B starting at (0x0 + 10$_{hex}$). Thus, Registers [8..10] values are now available. |
| 20: | wfbi | 0, 0, 0, 0, 40$_{hex}$; | Write data back to the frame buffer A from the output registers of column 1 into set 0, address 40$_{hex}$. Values for Circuits Registers [0..2]. |
| 21: | wfbi | 1, 0, 0, 0, 50$_{hex}$; | Write data back to the FB A from the output registers of column 2 into set 0, address 50$_{hex}$. Values for Circuits Registers [4..6]. |
| 22: | wfbi | 2, 0, 0, 0, 55$_{hex}$; | Write data back to the frame buffer A from the output registers of column 0 into set 0, address 55$_{hex}$. Values for Circuits Registers [8..10]. |
| 23: | dbcdc | r4, 10$_{hex}$, 0, 0, 0, 43$_{hex}$; | Double bank column 0 broadcast. Bank A starting at 43$_{hex}$. Bank B starting at (0x0 + 10$_{hex}$). The output calculated here is the xor operation of XOR$_0$, XOR$_7$, & XOR$_3$ the value is in cell 0 c 0. |
| 22: | wfbi | 0, 0, 0, 0, 60$_{hex}$; | Write data back to the FB A from the output registers of column 0 into set 0, address 60$_{hex}$. |
| 23: | dbcdc | r4, C$_{hex}$, 0, 0, 0, 60$_{hex}$; | Double bank column 2 broadcast. Bank A starting at 60$_{hex}$. Bank B starting at (0x0 + C$_{hex}$). Thus, Register$_3$ value is now calculated. |
| 24: | wfbi | 0, 0, 0, 0, 60$_{hex}$; | Write data back to the FB A from the output registers of column 0 into set 0, address 60$_{hex}$. |
| 25: | dbcdc | r4, 10$_{hex}$, 0, 0, 0, 30$_{hex}$; | Double bank column 2 broadcast Bank A starting at 30$_{hex}$. Bank B starting at (0x0 + 10$_{hex}$). Thus, Register$_7$ value is now calculated. |
| 26: | wfbi | 0, 0, 0, 0, 65$_{hex}$; | Write data back to the FB A from the output registers of column 0 into set 0, address 65$_{hex}$. |

Table 4. The TinyRISC code for the CRC CCITT-16 Algorithm.

| 0: | ldui | r1, 0x1; | R1 ← 10000hex. This is where DataIn Stored |
|---|---|---|---|
| 1: | ldfb | r1, 0, 0, 2 ; | FB ← 2 x 32 bits at set 0, bank A, address 0. |
| 2: | add | r0, r0, r0; | No-operation. |
| . | . | . | |
| 6: | ldui | r2, 0x1; | R2 ← 20000hex. This is where Circuit Registers Stored. |
| 7: | ldfb | r2, 1, 0, 2 ; | FB ← 2 x 32 bits at set 0, bank B, address 0. |
| 8: | add | r0, r0, r0; | No-operation. |
| . | . | . | |
| 12: | ldui | r3, 0x3; | R3 ← 30000hex. This is where the context word is stored in main memory. |
| 7: | ldctxt | r3, 0, 0, 0, 3; | Load 3 context words from main memory starting at the address stored in register 3 into plane 0, block 0 and starting at word 0. |
| 8: | add | r0, r0, r0; | NOP |
| . | . | . | |
| 11: | ldui | r4, 0x0; | R4 ← 00000hex. |
| 12: | dbcdc | r4, 0, 0, 0, 0, 0, 0; | Double bank column broadcast. It sends data from both banks address 0 in the frame buffer and broadcasts the context words column-wise. It triggers the RC array to start execution of column 0 by the context word of address 0 in the column block of context memory operating on data in set 0. Bank A starting at 0x0. Bank B starting at (0x0 + 0). |
| 13: | wfbi | 0, 0, 0, 0, 0x0; | Write data back to the frame buffer A from the output registers of column 0 into set 0, address 0. Results for XOR_i. |
| 14: | wfbi | 0, 0, 1, 0, 10hex; | Write data back to FB B from the output registers of column 0 into set 0, address 10hex. |
| 15: | sbcb | 0, 0, 0, 1, 0, 0hex; | Single bank column broadcast causing all the cells in the RC array column 0 to perform their operations specified by the second context word in context memory starting with data from frame buffer, set 0, bank A |
| 16: | sbcb | 0, 1, 0, 1, 0, 1hex; | It sends data from address 1hex in FB Column 1. |
| 17: | sbcb | 0, 2, 0, 1, 0, 2hex; | It sends data from address 2hex in FB Column 2. |
| 18: | sbcb | 0, 3, 0, 1, 0, 3hex; | It sends data from address 3hex in FB. |
| 19: | sbcb | 0, 4, 0, 1, 0, 4hex; | It sends data from address 4hex in FB. |
| 20: | sbcb | 0, 5, 0, 1, 0, 5hex; | It sends data from address 5hex in FB. |
| 21: | sbcb | 0, 6, 0, 1, 0, 6hex; | It sends data from address 6hex in FB, the new value of Register_14. |
| 22: | sbcb | 0, 7, 0, 1, 0, 7hex; | It sends data from address 7hex in FB, the new value of Register_15. |
| 23: | wfbi | 6, 0, 0, 0, 30hex; | Write data back to FB A from the output registers of column 6 into set 0, address 30hex. Value of Register_14. |
| 24: | wfbi | 7, 0, 0, 0, 31hex; | Write data back to the FB A from the output registers of column 7 into set 0, address 31hex. Value of Register_15. |
| 25: | dbcdc | r4, 13hex, 0, 0, 0, 2hex; | Double bank column broadcast. Bank A starting at 2hex. Bank B starting at 13hex. New values for Registers [8..13] are now available. |
| 26: | wfbi | 0, 0, 0, 0, 35hex; | Write data back to the FB A from the output registers of column 0 into set 0, address 35hex. |
| 27: | dbcdc | r4, 8hex, 0, 0, 0, 31hex; | Double bank column broadcast. It sends data from Bank A starting at 31hex. Bank B starting at 8hex. New value for Register_0. |
| 28: | wfbi | 0, 0, 0, 0, 45hex; | Write data back to the FB A from the output registers of column 0 into set 0, address 45hex. |
| 29: | dbcdc | r4, 14hex, 0, 0, 0, 0hex; | Double bank column broadcast. It sends data from Bank A starting at 0hex. Bank B starting at 14hex. New values for Register_6. |
| 30: | wfbi | 0, 0, 0, 0, 50hex; | Write data back to the FB A from the output registers of column 0 into set 0, address 50hex. |
| 31: | dbcdc | r4, 10hex, 0, 0, 0, 35hex; | Double bank column broadcast. It sends data from Bank A starting at 2hex. Bank B starting at 13hex. New values for Register_7. |
| 30: | wfbi | 0, 0, 0, 0, 55hex; | Write data back to the FB A from the output registers of column 0 into set 0, address 45hex. |

Table 5. The TinyRISC code for the CRC-16 Algorithm.

| Label | | | Description | Clocks | |
|---|---|---|---|---|---|
| | | | | 80486 | Pentium |
| | MOV | SP, D_Loc | SP ← Location of DataIn in memory. | 1T | 1T |
| | MOV | BP, IR_Loc | BP ← Location of Registers initial values in memory. | 1T | 1T |
| | MOV | DI, XORs_Loc | SP ← Loc of the resultant XORs in memory. | 1T | 1T |
| | MOV | BX, C_Value | BX ← Counter Value. | 1T | 1T |
| ZIP: | MOV | AL, [SP] | Get the vector element addressed by the SP register. | 1T | 1T |
| | MOV | BL, [BP] | Get the vector element addressed by the BP register. | 1T | 1T |
| | XOR | BL, AL | BL ← AL XOR BL. | 1T | 1T |
| | MOV | [DI], BL | | 1T | 1T |
| | INC | SP | Get next element. | 1T | 1T |
| | INC | BP | Get next element. | 1T | 1T |
| | INC | DI | Increment the destination. | 1T | 1T |
| | DEC | BX | Decrement the counter. | 1T | 1T |
| | JNZ | ZIP | Jump of not finished to label AA. | 3/1T | 1T |
| | MOV | AL, [DI ] | load the value of $XOR_0$ in AL. | 1T | 1T |
| | XOR | AL, [DI + 4] | XOR with $XOR_4$. | 1T | 1T |
| | MOV | $20000, AL | Store the result in memory location $20000_{hex}$ | 1T | 1T |
| | MOV | AL, [DI + 1] | load the value of $XOR_0$ in AL. | 1T | 1T |
| | XOR | AL, [DI + 5] | XOR with $XOR_5$. | 1T | 1T |
| | MOV | $20001, AL | Store the result in memory location $20001_{hex}$ | 1T | 1T |
| | MOV | AL, [DI + 2 ] | load the value of $XOR_2$ in AL. | 1T | 1T |
| | XOR | AL, [DI + 6] | XOR with $XOR_6$. | 1T | 1T |
| | MOV | $20002, AL | Store the result in memory location $20002_{hex}$ | 1T | 1T |
| | MOV | AL, [DI + 3] | load the value of $XOR_3$ in AL. | 1T | 1T |
| | XOR | AL, [DI + 7] | XOR with $XOR_7$. | 1T | 1T |
| | MOV | $20003, AL. | Store the result in memory location $20003_{hex}$ | 1T | 1T |
| | MOV | AL, [BP + 8] | Store the value of $Register_8$ in al. | 1T | 1T |
| | XOR | AL, $20000 | XOR the contents of $Register_8$ | 1T | 1T |
| | MOV | $30000, AL | Store the new value in $Register_0$. | 1T | 1T |
| | MOV | AL, [BP + 9] | Store the value of $Register_9$ in AL. | 1T | 1T |
| | XOR | AL, $20001 | XOR the contents of $Register_9$ | 1T | 1T |
| | MOV | $30001, AL | Store the new value in $Register_1$. | 1T | 1T |
| | MOV | AL, [BP + 10] | Store the value of $Register_{10}$ in AL. | 1T | 1T |
| | XOR | AL, $20002 | XOR the contents of $Register_{11}$ | 1T | 1T |
| | MOV | $30002, AL | Store the new value in $Register_2$. | 1T | 1T |
| | MOV | AL, [BP + 11] | Store the value of $Register_{11}$ in AL. | 1T | 1T |
| | XOR | AL, $20003 | XOR the contents of $Register_8$ | 1T | 1T |
| | XOR | AL, [DI] | XOR the contents with $XOR_0$ | 1T | 1T |
| | MOV | $30003, AL | Store the new value in $Register_3$. | 1T | 1T |
| | MOV | AL, [BP + 12] | Store the value of $Register_{12}$ in AL. | 1T | 1T |
| | XOR | AL, [DI + 1] | XOR with $XOR_1$. | 1T | 1T |
| | MOV | $30004, AL | Store the new value in $Register_4$. | 1T | 1T |
| | MOV | AL, [BP + 13] | Store the value of $Register_{13}$ in AL. | 1T | 1T |
| | XOR | AL, [DI + 2] | XOR with $XOR_2$. | 1T | 1T |
| | MOV | $30005, AL | Store the new value in $Register_5$. | 1T | 1T |
| | MOV | AL, [BP + 14] | Store the value of $Register_{14}$ in AL. | 1T | 1T |
| | XOR | AL, [DI + 3] | XOR with $XOR_3$. | 1T | 1T |
| | MOV | $30006, AL | Store the new value in $Register_6$. | 1T | 1T |
| | MOV | AL, [BP + 15] | Store the value of $Register_{15}$ in AL. | 1T | 1T |
| | XOR | AL, $20000 | XOR with data in mem(20000) | 1T | 1T |
| | MOV | $30007, AL | Store the new value in $Register_7$. | 1T | 1T |
| | MOV | AL, [DI] | Store the value of $XOR_0$ in AL. | 1T | 1T |
| | XOR | AL, $20001 | XOR with data in mem(20001) | 1T | 1T |
| | MOV | $30008, AL | Store the new value in $Register_8$. | 1T | 1T |
| | MOV | AL, [DI + 1] | Store the value of $XOR_1$ in AL. | 1T | 1T |
| | XOR | AL, $20002 | XOR with data in mem(20002) | 1T | 1T |
| | MOV | $30009, AL | Store the new value in $Register_9$. | 1T | 1T |
| | MOV | AL, [BP + 2] | Store the value of $XOR_2$ in AL. | 1T | 1T |
| | XOR | AL, $20003 | XOR with data in mem(20003) | 1T | 1T |
| | MOV | $3000A, AL | Store the new value in $Register_{10}$. | 1T | 1T |
| | MOV | AL, [BP +3] | Store the value of $XOR_3$ in AL. | 1T | 1T |
| | MOV | $3000B, AL | Store the new value in $Register_{11}$. | 1T | 1T |
| | MOV | $3000C, $20000 | Store the new value in $Register_{12}$. | 1T | 1T |
| | MOV | $3000D, $20001 | Store the new value in $Register_{13}$. | 1T | 1T |
| | MOV | $3000E, $20002 | Store the new value in $Register_{14}$. | 1T | 1T |
| | MOV | $3000F, $20003 | Store the new value in $Register_{15}$. | 1T | 1T |
| **Findings** | | | **Total Clocks** | **142T** | **128T** |
| | | | **Frequency** | **100MHz** | **133MHz** |
| | | | **Total Time** | **1.42µsec** | **0.96µsec** |
| | | | **Number of Bits** | **8** | **8** |
| | | | **Bits per Cycles** | **0.056** | **0.0625** |
| | | | **Mega Bits per Second (Mbps)** | **5.6** | **8.3** |
| | | | **Cycles per Bits** | **17.86** | **16** |

Table 6. The Intel code for the CRC-CCITT-16 Algorithm.

| | | | | Clocks | |
|---|---|---|---|---|---|
| **Label** | | | **Description** | **80486** | **Pentium** |
| | MOV | SP, D_Loc | SP ← Location of DataIn in memory. | 1T | 1T |

| | MOV | BP, IR_Loc | BP ← Location of Registers initial values in memory. | 1T | 1T |
|---|---|---|---|---|---|
| | MOV | DI, XOR$_8$_Loc | SP ← Loc of the resultant XORs in memory. | 1T | 1T |
| | MOV | BX, C_Value | BX ← Counter Value. | 1T | 1T |
| **ZIP:** | MOV | AL, [SP] | Get the vector element addressed by the SP register. | 1T | 1T |
| | MOV | BL, [BP] | Get the vector element addressed by the BP register. | 1T | 1T |
| | XOR | BL, AL | BL ← AL XOR BL. | 1T | 1T |
| | MOV | [DI], BL | | 1T | 1T |
| | INC | SP | Get next element. | 1T | 1T |
| | INC | BP | Get next element. | 1T | 1T |
| | INC | DI | Increment the destination. | 1T | 1T |
| | DEC | BX | Decrement the counter. | 1T | 1T |
| | JNZ | *ZIP* | Jump of not finished to label AA. | 3/1T | 1T |
| | MOV | BX, C_Value | BX ← Counter Value. | 1T | 1T |
| **FUNNEL:** | MOV | AL, $0 | Assign zero to AL. | 1T | 1T |
| | MOV | BL, [DI] | Get the vector element addressed by the BP register. | 1T | 1T |
| | XOR | AL, BL | AL ← AL XOR BL. | 1T | 1T |
| | INC | DI | Increment the destination. | 1T | 1T |
| | DEC | BX | Decrement the counter. | 1T | 1T |
| | JNZ | *FUNNEL* | Jump of not finished to label AA. | 3/1T | 1T |
| | MOV | $20000, BL | Store the result of funneling in memory location 20000$_{hex}$. | 1T | 1T |
| | MOV | AL, [BP + 8] | Store the value of Register$_8$ in AL. | 1T | 1T |
| | XOR | AL, $20000 | XOR the contents of Register$_8$ and the result of Funneling | 1T | 1T |
| | MOV | $30000, AL | Store the new value in Register$_0$. | 1T | 1T |
| | MOV | $30001, [BP + 9] | Store the value found in Register$_9$ in the new location of Register$_1$. | 1T | 1T |
| | MOV | $30002, [BP + 10] | Register$_2$←Register$_{10}$. | 1T | 1T |
| | MOV | $30003, [BP + 11] | Register$_3$←Register$_{11}$. | 1T | 1T |
| | MOV | $30004, [BP + 12] | Register$_4$←Register$_{12}$. | 1T | 1T |
| | MOV | $30005, [BP + 13] | Register$_5$ ←Register$_{13}$. | 1T | 1T |
| | MOV | AL, [BP + 14] | Store the value of Register$_{14}$ in AL. | 1T | 1T |
| | XOR | AL, [DI] | XOR the contents of Register$_{14}$ with XOR$_0$. | 1T | 1T |
| | MOV | $30006, AL | Store the new value in Register$_6$. | 1T | 1T |
| | MOV | AL, [BP + 15] | Store the value of Register$_{15}$ in AL. | 1T | 1T |
| | XOR | AL, [DI] | XOR the contents of Register$_{14}$ with XOR$_0$. | 1T | 1T |
| | XOR | AL, [DI + 1] | XOR the contents with XOR$_1$. | 1T | 1T |
| | MOV | $30007, AL | Store the new value in Register$_7$. | 1T | 1T |
| | MOV | AL, [DI + 2] | Store the value of XOR$_2$ in AL. | 1T | 1T |
| | XOR | AL, [DI + 3] | XOR with XOR$_3$. | 1T | 1T |
| | MOV | $30008, AL | Store the new value in Register$_8$. | 1T | 1T |
| | MOV | AL, [DI + 3] | Store the value of XOR$_3$ in AL. | 1T | 1T |
| | XOR | AL, [DI + 4] | XOR with XOR$_4$. | 1T | 1T |
| | MOV | $30009, AL | Store the new value in Register$_9$. | 1T | 1T |
| | MOV | AL, [DI + 4] | Store the value of XOR$_4$ in AL. | 1T | 1T |
| | XOR | AL, [DI + 5] | XOR with XOR$_5$. | 1T | 1T |
| | MOV | $3000A, AL | Store the new value in Register$_{10}$. | 1T | 1T |
| | MOV | AL, [DI + 5] | Store the value of XOR$_5$ in AL. | 1T | 1T |
| | XOR | AL, [DI + 6] | XOR with XOR$_6$. | 1T | 1T |
| | MOV | $3000B, AL | Store the new value in Register$_{11}$. | 1T | 1T |
| | MOV | AL, [DI + 6] | Store the value of XOR$_6$ in AL. | 1T | 1T |
| | XOR | AL, [DI + 7] | XOR with XOR$_7$. | 1T | 1T |
| | MOV | $3000C, AL | Store the new value in Register$_{12}$. | 1T | 1T |
| | MOV | AL, [DI + 7] | Store the value of XOR$_7$ in AL. | 1T | 1T |
| | XOR | AL, [DI + 8] | XOR with XOR$_8$. | 1T | 1T |
| | MOV | $3000D, AL | Store the new value in Register$_{13}$. | 1T | 1T |
| | MOV | AL, [DI + 8] | Store the value of XOR$_8$ in AL. | 1T | 1T |
| | XOR | AL, $20000 | XOR the contents with the result of Funneling | 1T | 1T |
| | MOV | $3000E, AL | Store the new value in Register$_{14}$. | 1T | 1T |
| | MOV | $3000F, $20000 | Store the new value in Register$_{15}$. | 1T | 1T |
| **Findings** | | **Total Clocks** | | **190T** | **162T** |
| | | **Frequency** | | **100MHz** | **133MHz** |
| | | **Total Time** | | **1.9μsec** | **1.22μsec** |
| | | **Number of Bits** | | **8** | **8** |
| | | **Bits per Cycles** | | **0.042** | **0.049** |
| | | **Mega Bits per Second (Mbps)** | | **4.2** | **6.56** |
| | | **Cycles per Bits** | | **23.75** | **20.25** |

Table 7. The Intel code for the CRC-16 Algorithm.

| **Algorithm** | System | N# of Cycles | Speedup | Time in Micro Sec. | Bits per Cycle | Mega Bits Per Second | Cycles per Bits |
|---|---|---|---|---|---|---|---|
| | | | | | | | |
| **CRC-CCITT-16 parallel algorithm for one channel** | M1 | 30 | | 0.3 | 0.267 | 26.67 | 3.75 |
| | Pentium | 128 | 4.26 | 0.96 | 0.0625 | 8.3 | 16 |
| | 80486 | 142 | 4.73 | 1.42 | 0.056 | 5.6 | 17.86 |
| **CRC-16 parallel algorithm for one channel** | M1 | 26 | | 0.26 | 0.307 | 30.76 | 3.25 |
| | Pentium | 162 | 6.23 | 1.22 | 0.049 | 6.56 | 20.25 |
| | 80486 | 190 | 7.3 | 1.9 | 0.042 | 4.2 | 23.75 |
| **CRC-CCITT-16 parallel algorithm for 8-channel** | M1 | 30 | | 0.3 | 2.13 | 213.13 | 0.46 |
| | Pentium | 1024 | 34.13 | 7.69 | 0.0625 | 8.32 | 16 |
| | 80486 | 1136 | 37.86 | 11.36 | 0.056 | 5.63 | 17.75 |
| **CRC-16 parallel algorithms for 8-channels** | M1 | 26 | | 0.26 | 2.46 | 246.15 | 0.41 |
| | Pentium | 1296 | 49.84 | 9.74 | 0.05 | 6.57 | 20.25 |
| | 80486 | 1520 | 58.46 | 15.2 | 0.042 | 4.21 | 23.75 |

Table 8. Comparison with the Intel systems.

| Algorithm | System | N# of Cycles | Speedup of the RC-1000 over the M1 |
|---|---|---|---|
| **CRC-CCITT-16 parallel algorithm** | M1 | 30 | |
| | RC-1000 | 8 | 3.75 |
| **CRC-16 parallel algorithm** | M1 | 26 | |
| | RC-1000 | 17 | 1.53 |

Table 9. Comparisons with RC-1000 FPGA.